\documentclass[twocolumn,showpacs,preprintnumbers,amsmath,amssymb]{revtex4}
\usepackage{graphicx}
\usepackage{dcolumn}
\usepackage{bm}
\usepackage[english]{babel}

\begin{document}

%

\preprint{APS/123-QED}

\title{On chaotic behavior of gravitating stellar shells.}

 \author{M.V. Barkov}
  \email{ barmv@sai.msu.ru}
 \author{G.S. Bisnovatyi-Kogan}
 \email{ gkogan@mx.iki.rssi.ru}
 \author{A.I. Neishtadt}
 \email{aneishta@mx.iki.rssi.ru}
 \affiliation{Space Research Institute, 84/32 Profsoyuznaya Str , Moscow,
Russia, 117997}

 \author{V.A. Belinski}
 \affiliation{National Institute of Nuclear Physics (INFN) and
International Center of Relativistic Astrophysics (ICRA), Dip. di
Fisica - Universita` degli Studi di Roma "La Sapienza" P.le Aldo
Moro, 5 - 00185 Roma, Italy}
 \email{belinski@icra.it}

\date{\today}

\begin{abstract}
Motion of two gravitating spherical stellar shells around a
massive central body is considered. Each shell consists of point
particles with the same specific angular momenta and energies. In
the case when one can neglect the influence of gravitation of one
("light") shell onto another ("heavy") shell ("restricted
problem") the structure of the phase space is described. The
scaling laws for the measure of the domain of chaotic motion and
for the minimal energy of the light shell sufficient for its
escape to infinity are obtained.
\end{abstract}


\maketitle

{\bf Chaotic motion of a system can result in crossing a border in
the phase space separating the area where disintegration of the
system is possible. In the present paper this phenomenon is
studied in a system of two gravitating stellar shells oscillating
around a massive central body. Similar phenomena take place in
planetary and molecular dynamics.}

\section{Introduction}

Dynamical processes around supermassive black holes (SBH) in
quasars, blazars and active galactic nuclei  are characterized by
violent phenomena, leading to formation of jets and other
outbursts. Here we consider a model problem for a shell outburst
from a SBH surrounded by a dense massive stellar cluster due to a
pure gravitational interaction of shells oscillating around SBH.

Investigation of spherical stellar clusters using shell
approximation was started by H\'enon \cite{he1}. In this
approximation a stellar cluster is considered as a collection of
spherical stellar shells. Each shell consists of stars with the
same specific  angular momenta and energies. In the process of
motion the  angular momenta are conserved while  energies are
changing, yet taking the same values for all the stars in one
shell. The shell itself stay spherical. Shell approximation has
been successfully applied for investigation of stability
\cite{he2}, violent relaxation and collapse \cite{Ly67,he1,got},
leading to formation of a stationary cluster. Investigation of the
evolution of a spherical stellar cluster taking into account
various physical processes was done on the base of the shell model
in the classical series of papers of L. Spitzer and his coauthors
\cite{s1,s2,s3,s4,s5,s6,s7,s8}, see also \cite{vog}.

Numerical calculations of a collapse of stellar clusters in
theshell approximation \cite{x8,x4,x5} have shown, that even if
all shells are  initially gravitationally bound, after a number of
passages through each other some shells obtain sufficient energy
to be thrown to infinity.

Here  we consider a simplified problem of  motion of two massive
spherical shells, each consisting of stars with the same specific
angular momentum and energy, around SBH.

Development   of   chaos   during   the  motion  of  two
gravitating  intersecting  shells  was  found first by Miller and
Youngkins  \cite{x7} in an oversimplified case of purely radial
motion and reflecting inner boundary. In \cite{bbb_mnras,bbb_jetp}
chaotic behavior was found in  a more realistic model considering
stars with the same specific angular momentum and energy dispersed
isotropically over the spherical  shell,  and  each star moving
along its ballistic trajectory in the  gravitational field of the
central body and shells. In the current paper we consider the same
model mainly in the case  when one can neglect the influence of
one ("light") shell onto the motion of another ("heavy") shell. In
this simplified case we describe a structure of the phase space
and obtain the scaling laws for the measure of the domain of
chaotic motion and for  the minimal energy of the light shell
sufficient for its escape to infinity.

 We follow the approach which was used in \cite{pet} for analysis
of chaotic dynamics and escape to infinity in the planar
restricted circular three body problem. This approach consists of
constructing asymptotic formulas for Poincar\'e return map  and
identification of regions of regular and chaotic dynamics in the
phase space of this map. Region of regular dynamics is the region
where Kolmogorov-Arnold-Moser (KAM) theory (see, e.g.
\cite{arn,akn}) guarantees existence of many invariant curves of
the map. These curves are barriers preventing the escape to
infinity. Region of chaotic motion is the region where the phase
expansion criterion (see, e.g. \cite{zasl}) is satisfied. In this
region chaotic diffusion allows the "light"{} shell to gain
positive energy and then escape to infinity. One special property
of the system under consideration is that for an open set of the
problem's parameters, in the phase space there is a region where
the Poincar\'e return map is continuous but not smooth (it has a
singularity of the square root kind). In this region neither
KAM-theory nor phase expansion criterion are applicable. Similar
situation, though with a map that is discontinuous in a certain
region in the phase space occur in the problem of motion of three
gravitating parallel impenetrable sheets \cite{physd}. For general
analysis of maps with discontinuities see \cite{ashw} and
references there.

The dynamics of a system of  many gravitating spherical shells was
studied from the viewpoint of statistical mechanics in \cite{pre1,
pre2,pre3} where, however, the escape to infinity was not studied
because a bounded system was considered. An extensive study of
escape to infinity in the rectilinear three body problem was
performed in \cite{chaos}.

\section{ A restricted problem.}

Consider the motion of two stellar shells with masses $m_1$ and
$m_2$ around the central body of mass $M$. Let $J_1$ and $J_2$ be
specific  angular  momenta of shells. Consider the case when $m_2
\ll m_1$, so that one can neglect the influence of the second
shell onto the motion of the first shell. We call this problem a
restricted problem by the analogy with the restricted three body
problem in celestial mechanics (the latter problem describes the
motion of an asteroid under the action of Sun and Jupiter).

In what follows the first shell will be called the heavy one, and
the second shell will be called the light one.

In the restricted problem, motion of the heavy shell is described
by Hamiltonian function
\begin{equation}
 \label{ch1}
 H_1=\frac{p^2_1}{2}+V_1, \quad V_1=-\frac{G\left(M+m_1/2\right)}{r_1}
 +\frac{J^2_1}{2r^2_1}.
\end{equation}
Here $p_1$ is the momentum canonically conjugated to the shell
radius $r_1$. Therefore, the change of $r_1$ in time is the same
as in the Kepler problem for a satellite with mass 1 and angular
momentum $J_1$ and a central body of mass $(M+m_1 / 2)$. For
$H_1=h_1<0$ the behavior of $r_1$ is described by usual formulas
of the elliptic Keplerian motion: $r_1=r_1(l_1,h_1,J_1)$; here
$l_1$ is the mean anomaly of the satellite, i.e. the angular
variable for which
\begin{equation}
 \label{ch2}
 \dot l_1 = \frac{\sqrt{\mu_1}}{a_1^{3/2}},\quad
 \mu_1 = G\left(M+m_1/2\right), \quad
 a_1=-\frac{\mu_1}{2h_1}.
\end{equation}
Motion of the light shell is described by Hamiltonian
\begin{eqnarray}
 \label{ch3}
  H_2&=&\frac{p_2^2}{2}+V_2, \nonumber \\
 V_2&= &
 \left\{
    \begin{array}{lll}
        V_0,& \mbox{if} & r_2\leq r_1,\\
        V_0-{Gm_1}/{r_2},&\mbox{if}& r_2 \ge r_1,
    \end{array}
 \right.
\end{eqnarray}
where
 $$
 V_0=-\frac{GM}{r_2}+\frac{J_2^2}{2r_2^2}.
 $$
Here $p_2$ is the momentum canonically conjugated to radius $r_2$
of the light shell. In the Hamiltonian $H_2$ one should consider
$r_1$ as a given  function of time. Therefore, in the restricted
problem motion of the light shell is described by a Hamiltonian
system with one and a half  degrees of freedom (one degree of
freedom plus explicit dependence of the Hamiltonian on time). {
This model applies just as well to the case where the light
massless shell is replaced by a massless point. }

\section{\label{sec4}  Poincar\'e return map for the restricted problem.}

Consider the motion of the light shell. At the time moments $t$,
when $r_2$ has a local minima on the trajectory, we mark on the
cylinder $\Phi = \left\{ \left(h,\phi\right): -\infty <h<\infty,\;
\phi \mod 2\pi\right\}$ the values of the light shell Hamiltonian
$H_2$ and of the heavy shell mean anomaly $l_1 : (H_2,l_1)\equiv
(h,\phi)$. Thus we obtain on $\Phi$ a sequence of points $\ldots,$
$(h^{s},\phi^{s}),$ $\ldots,$ $(h^{-1},\phi^{-1}),$
$(h^{0},\phi^{0}),$$(h^{1},\phi^{1}),$ $\ldots ,$
$(h^{k},\phi^{k}),$ $\ldots  \; $. This sequence terminates at the
right if the shell escapes to infinity. It can be terminated at
the left as well, but the terms of this sequence with numbers
smaller than certain negative number do not have physical meaning.
The points of the constructed sequence are mapped to each other
under the action of Poincar\'e return map defined as follows.
Consider the motion of the light shell that starts at time moment
$t=\tilde{t}$ with $p_2 = 0$. Denote initial values of $H_2$ and
$l_1$ as $h$ and $\phi$. The initial value of $r_2$ satisfies the
equation $H_2=h$. Let this initial value $r_2$ be the smallest of
two roots of equation $H_2=h$. Therefore, at the time moment
$\tilde{t}$ the value of $r_2$ on the trajectory has a local
minimum. Denote $t'$ as the first time moment after $\tilde{t}$
such that at this time moment again $p_2=0 $ and $r_2$ on the
trajectory has a minimum, provided that such a time moment exists.
Denote as $h'$ and $\phi' $ values of $H_2$ and $l_1$
corresponding to $t=t'$. By definition, Poincar\'e return map
$\Pi$ is given by the formula $\Pi(h,\phi)=(h', \phi')$. Then
$\left(h^{(k+1)},\phi^{(k+1)}\right)=\Pi\left(h^{(k)},\phi^{(k)}\right)$.
The map $\Pi$ is defined on the part of the cylinder $\Phi$ where
values of $(h,\phi)$ allow the shell to make at least one complete
oscillation. The map $\Pi$ preserves the area $dhd\phi$ on $\Phi$;
this follows from preservation of the phase volume by a
Hamiltonian system (see, e.g. Refs.~\cite{arn,akn}).

Together with map $\Pi$ we will use modified return map
$\hat{\Pi}$ defined as follows. Let $\tilde{t},t',{}\phi,\phi'$ be
the same as before. Denote by $\hat{h}$ the value of $H_2$ at the
last preceding to $\tilde{t}$ time moment when $r_2$ has a
maximum. Denote by $\hat{h}'$ the value of $H_2$ at the first time
moment after $\tilde{t}$ when $r_2$ has a maximum. Define map
$\hat{\Pi}$ by the formula
$\hat{\Pi}\left(\hat{h},\phi\right)=\left(\hat{h}',\phi'\right)$.
The map $\hat{\Pi}$ is defined on the part of the cylinder
$\hat{\Phi} = \left\{ \left(\hat{h},\phi\right): -\infty
<\hat{h}<\infty,\; \phi \mod 2\pi\right\}$ where $\hat{h}<0$. The
map $\hat{\Pi}$ preserves the area $\left(\partial h/\partial
\hat{h}\right)_{\phi}d\hat{h} d\phi$ on $\hat{\Phi}$. For a given
$\left(\hat{h}^{l},\phi^{l}\right)$ the map $\hat{\Pi}$ generates
on $\hat{\Phi}$ the sequence of points  $\ldots,$
$(\hat{h}^{s},\phi^{s}),$ $\ldots ,$ $(\hat{h}^{-1},\phi^{-1}),$
$(\hat{h}^{0},\phi^{0}),$$(\hat{h}^{1},\phi^{1}),$ $\ldots ,$
$(\hat{h}^{k},\phi^{k}),$ $\ldots $ such that $\left(
\hat{h}^{(k+1)} , \phi^{(k+1)}\right) =
\hat{\Pi}\left(\hat{h}^{(k)} , \phi^{(k)}\right)$. At small values
of $\hat{h}$ the time interval between $\tilde{t}$ and $t'$ is
determined in the main approximation by the value $\hat{h}'$:
$t'-\tilde{t}=2\pi \mu_2\left/\left(-2\hat{h}'\right)^{3/2}\right.
, \; \mu_2=G\left(M+m_1\right)$. This is the reason why for mall
$\hat{h}$ map $\hat{\Pi} $ is more convenient for the analysis
than $\Pi$. In what follows we omit for brevity the symbol
"{}$\hat{}\;$"{} over $\hat{h},\hat{\Phi}$.

In what follows we consider the case when the heavy shell's mass
is much smaller than the mass of the central body:
$m_1=\varepsilon M,\; 0<\varepsilon\ll 1$. In this case map
$\hat{\Pi}$ has the form
\begin{equation}
 \label{ch4}
 h'=h+\varepsilon f\left(h,\phi,\varepsilon\right),
\end{equation}
\begin{equation}
 \label{ch5}
 \phi'=\phi+2\pi \frac{\mu_2}{\mu_1}\left(\frac{h_1}{h'}\right)^{3/2}
 +\varepsilon g\left(h',\phi',\varepsilon\right).
\end{equation}
Functions $f,g$ can be extended as continuous functions up to the
argument value $h=0$. The value $\varepsilon f\left( h,\phi,
\varepsilon \right)$ is the sum of jumps of the light shell's
potential energy due to passage of shells through each other
between two consecutive time moments when the light shell radius
takes maximal values. To find of $\varepsilon f\left( h,\phi, 0
\right)$, one should calculate the sum of the energy jumps under
the assumption that motion of the light shell is not affected by
the heavy shell (i.e. this motion is Keplerian one with total
energy $h$).

Depending on the properties of the shell motion at
$\varepsilon=0$, one can define three main types of $h$ values.

\subsubsection{Type A.}

At all initial phases $\phi$ of the heavy shell motion, the light
shell on the period of its motion passes through the heavy shell
at non-zero relative velocity. (If $h=0$, i.e. motion of the shell
stars is parabolic, then in this and following definitions one
should consider as a period of the light shell motion the whole
infinite time interval between arrival of the shell from infinity
and its departure to infinity.) In this case during a period of
the light shell motion the  number of passages of shells through
each other is the same for any initial phase of the heavy shell
motion. Therefore, the number of terms in the sum for $ f \left( h
, \phi , 0 \right)$ is the same for any $\phi$. At small enough
$\varepsilon$, the sum for $ f \left( h , \phi , \varepsilon
\right)$ has the same number of the terms. Therefore, function $ f
\left( h , \phi , \varepsilon \right)$ is analytic in  $ \left( h
, \phi \right)$, and it can be continued analytically in some
neighborhood of a circle around $h=0$.

\subsubsection{Type B.}

There exists certain initial phase $\phi_*$ of the heavy shell
motion such that at some time moment the shells superimpose with
zero relative velocity and non-zero relative acceleration, and
absolute velocities of both shells at this moment are different
from zero. At this moment trajectories of the shells have
quadratic tangency. When $\phi$ deviates from $\phi_*$ in one
direction, the superimposing  of the shells disappears. When
$\phi$ deviates from $\phi_*$ in another direction, this
superimposing decomposes into two close shell crossings (say,
light shell surpasses the heavy  one and then almost immediately
the heavy shell surpasses the light one). In this latter case two
new terms appear in the sum for  $ f \left( h , \phi , 0 \right)$;
these terms are close by absolute values, and they have different
signs. The sum of these terms is of order $\sqrt{|\phi-\phi_*|}$.
Therefore, in this case function  $ f \left( h , \phi , 0 \right)$
has at $\phi=\phi_*$ the singularity of type $\sqrt{\phi-\phi_*}$
or $\sqrt{\phi_*-\phi}$. At small variations of $h$ this
singularity persists. Function  $ f \left( h , \phi , \varepsilon
\right)$ has the same singularity at small $\varepsilon$. Value
$r_2$ that corresponds to time moment of tangency of shell
trajectories is uniquely defined. Indeed, at the moment of
tangency there should be $r_1=r_2,\; h_1+J_1^2/2r_1^2 =
h+J_2^2/2r_2^2$ (because velocities of the shells are equal to
each other) and $J_1^2/2r_1^3 \neq J_2^2/2r_2^3$. These conditions
define a unique value of  $r_2$.

\subsubsection{Type C.}

For any initial phase $\phi$ of the heavy shell motion the shells
never superimpose. Then at small enough $\varepsilon$ motion of
the light shell is also Keplerian.

\bigskip

There is also exceptional intermediate type of $h$ for which the
shells can meet with zero velocities (i.e. one of two extremal
values of radius of one shell coincide with an extremal value of
radius of another shell), but with nonzero relative acceleration.
Another exceptional case is $h=h_1$ and $J_1=J_2$. In this case
for some initial phase $\phi$ of the heavy shell the radii of two
shells coincide all the time.

\section{Dynamics of shells in the restricted problem.}

\begin{figure*}
\includegraphics[width=.90\textwidth]{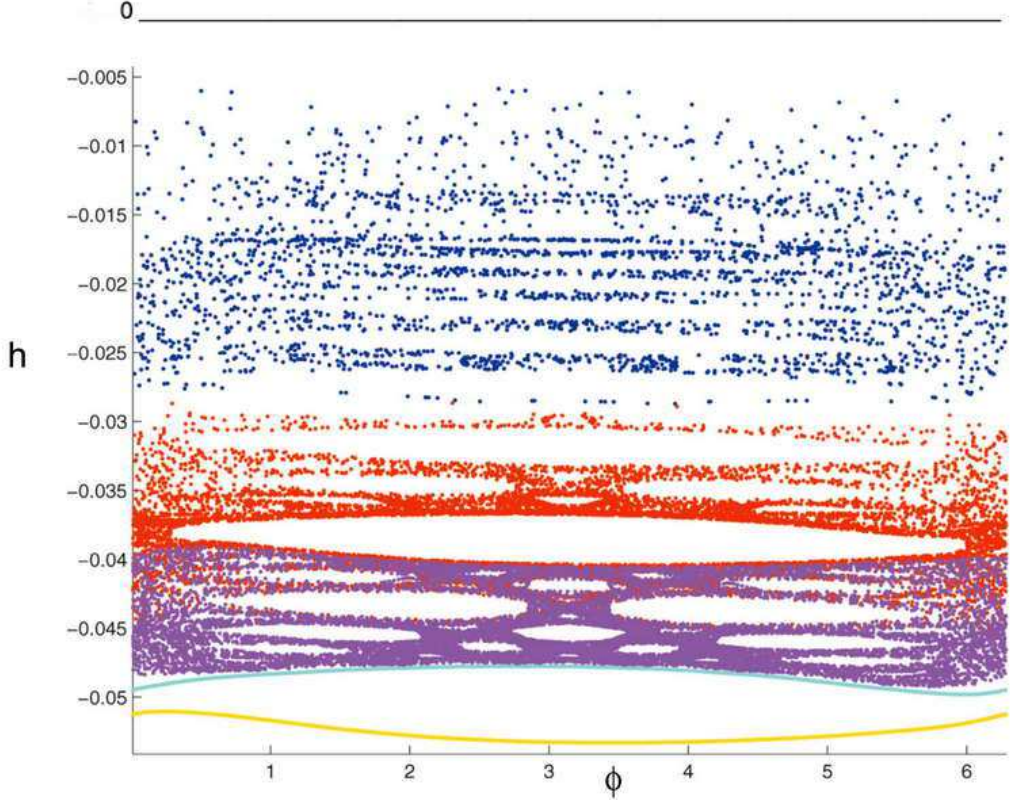}
\caption{\label{fig1} Trajectories of five initial points (3
chaotic, 2 regular)  under the action of Poincar\'e return map in
the domain  where  lines $h=$ const are of the type A.
$m_1/M=0.005$,
 $J_1=1.46\sqrt{2}$, $J_2=1.4\sqrt{2}$.}
\end{figure*}

\begin{figure*}
\includegraphics[width=.90\textwidth]{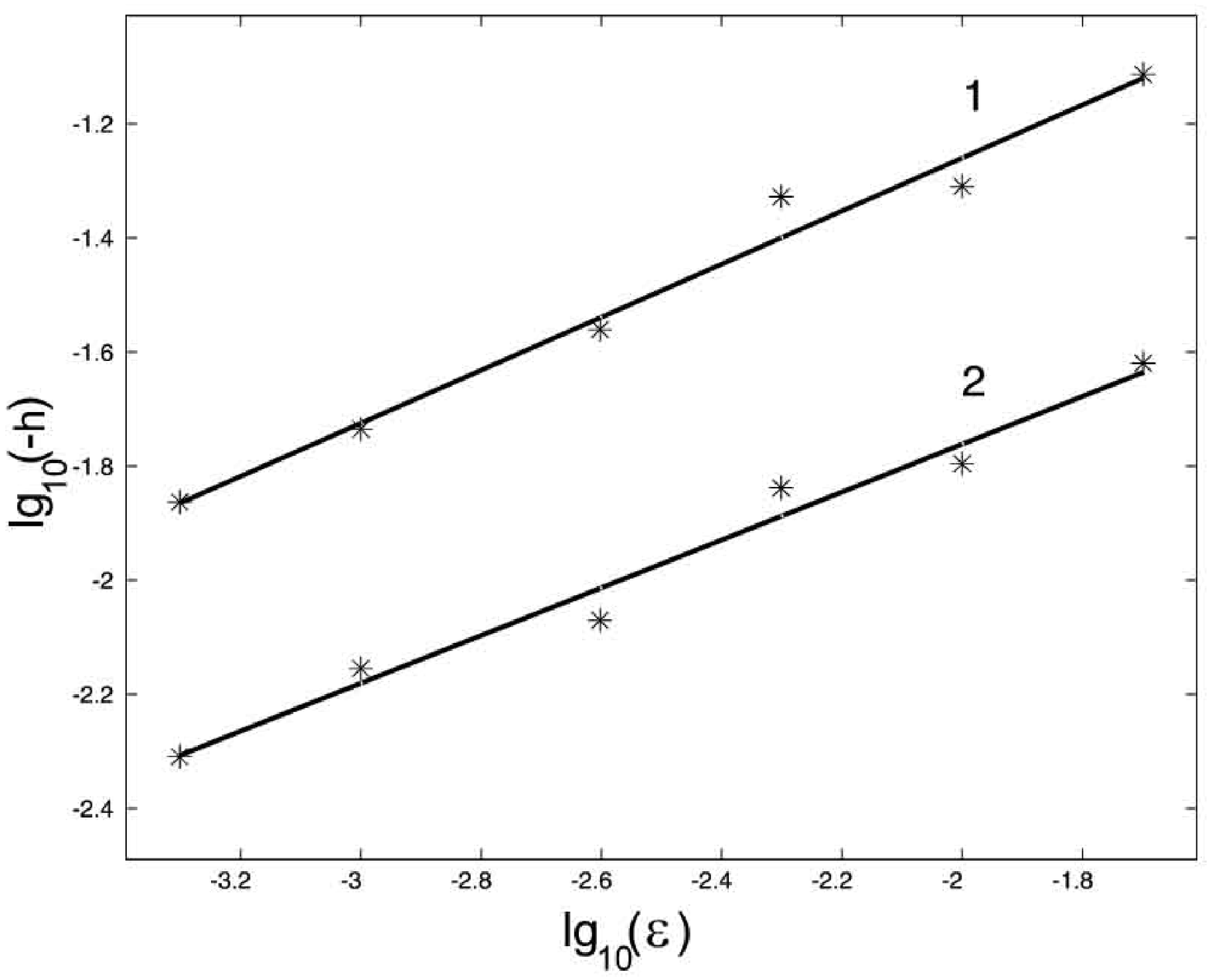}
\caption{\label{fig2} Maximal values of $h$ for the invariant
curves enclosing cylinder $\Phi$ and minimal value of $h$ such
that there are no visible islands at bigger values of $h$. The
line $h=0$ is of the type A.}
\end{figure*}

Let us consider first the case when the value $h=0$ is of type A.
Consider in cylinder $\Phi$ the annulus $\alpha\leq h\leq 0$ such
that all values $h$ on this annulus are of the type A (here
$\alpha$ is a constant, $\alpha<0$ does not depend on
$\varepsilon$). In this annulus the map $\hat{\Pi}$ is analogous
to "Kepler map"{}, which was introduced in \cite{pet} . Let us
repeat for map $\hat{\Pi}$ the analysis from  \cite{pet} (see also
\cite{pet16} ). The trajectories of several initial points under
the action of map $\hat{\Pi}$ for this case are shown in {Fig.
\ref{fig1}}. The properties of the map to large extent are
determined by the phase expansion coefficient
$\left(\partial\phi'/
\partial\phi\right)_{h}$. {  In our case relations (\ref{ch4}),
(\ref{ch5}) imply that }
\begin{equation}
 \label{ch6}
 \left(\frac{\partial\phi'}{ \partial\phi}\right)
 \sim
 \varepsilon  \left( \frac{h_1}{h}\right)^{5/2}.
\end{equation}
In the region where $\left(\partial\phi'/ \partial\phi\right)\ll
1$, i.e. $h<-K_1 h_1 \varepsilon^{2/5}$ with a big enough positive
constant $K_1$, map $\hat{\Pi}$ satisfies the assumptions of
KAM-theory. According to KAM theory, this region is filled up to a
residual set of small measure by invariant curves of map
$\hat{\Pi}$ { that enclose cylinder $\Phi$ (i.e. values of $\phi$
along any of these curves cover all segment $[0,2\pi]\; mod\;
2\pi$) and are close to curves $h=$ const. The residual set of
small measure is filled up to another residual set of much smaller
measure by invariant curves that do not enclose cylinder $\Phi$
(i.e. values of $\phi$ along each such curve cover only a part of
segment $[0,2\pi]\; mod\; 2\pi$).} These invariant curves are
organized into "islands"{} corresponding to different resonances.
For phase points on invariant curves the motion of the light shell
is quasi-periodic (in particular case, periodic). For a phase
point not on invariant curves the motion of the light shell can be
chaotic. However, because the trajectory of the phase point on
$\Phi$ remains locked between invariant curves, the light shell
can not acquire the positive energy and escape to infinity. The
invariant curves with largest $h$ value that can be constructed by
KAM theory have $h\sim-\varepsilon^{2/5}$, the variation of $h$
along any such curve is also of the order of $\varepsilon^{2/5}$.
Fig. \ref{fig2} represents results of numerical experiment in
which for several values of $\varepsilon$ we looked for the
invariant curve with the maximal value of $h$ enclosing cylinder
$\Phi$. The least mean square fit gives the scaling $h\sim -
\varepsilon^{0.46}$.

In the region where $\left(\partial\phi'/ \partial\phi\right)\gg
1$ small changes  in $\phi$ lead to large changes in $\phi'$.
Therefor, values of $\phi$ and $\phi'$ can be considered as
statistically independent according to to phase expansion
criterion. This is the region where $-K_2^{-1}h_1
\varepsilon^{2/5} < h < 0,$ with $K_2$ being a large enough
positive constant. In this region, variations of the light shell
energy $h$ during one iteration of map $\hat{\Pi}$ can be treated
as independent random values. As a result the changes of the
energy along the phase point trajectory are of diffusional type.
Apparently, as a result of this diffusion almost all phase points
in the region under consideration acquire positive energy, i.e.
light shell escape to infinity. Numerical experiment (see Fig.
\ref{fig2}, straight line 2) gives for the lower boundary of this
domain the scaling $h\sim - \varepsilon^{0.42}$. The lower
boundary of this domain was defined by the condition that there
are no visible islands in the numerical experiment above this
boundary.

In the intermediate region $-K_1 h_1 \varepsilon^{2/5} < h<
-K_2^{-1} h_1 \varepsilon^{2/5}$, there are both regular and
chaotic trajectories, and apparently the measures of sets of
corresponding phase points are of the same order
$\varepsilon^{2/5}$.

In any annulus $\gamma \leq h \leq \beta <0 $ with values $h$ of
the type A the dynamics is described by KAM theory (here $\beta,
\gamma$ are constants, i.e. values independent of $\varepsilon ,
\gamma< \beta$). Such an annulus is filled up to residual set of
measure $O\left( \sqrt{\varepsilon} \right)$ by invariant curves
enclosing cylinder $\Phi$; these curves are $O\left( \sqrt{
\varepsilon } \right)$-close to circles $h=$ const. The residual
set of measure $O\left( \sqrt{ \varepsilon } \right)$ is filled up
to residual set of much smaller measure, apparently $O\left(
\exp(-\mbox{const}/\sqrt{ \varepsilon }) \right)$, by invariant
curves that do not enclose cylinder $\Phi$ \cite{akn}. These
curves are organized into resonant "islands"{}. In this annulus,
therefore, chaotic motion can occupy only measure $O\left(
\exp(-\mbox{const}/\sqrt{ \varepsilon }) \right)$.

Let us now consider the case when value $h=0$ is of the type B.
Consider in cylinder $\Phi$ the annulus $\alpha\leq h \leq 0$ such
that all values h for this annulus are of type B (here $\alpha$ is
a constant, $\alpha$ does not depend on $\varepsilon$). In this
annulus map $\hat{\Pi}$ is continuous, but it is not smooth.
Singularities of $\Pi$ are described in Sec. \ref{sec4}.
KAM-theory is not applicable to such a map. Trajectories of
several initial points under the action of $\hat{\Pi}$ in this
case are shown in {Fig. 3}. Instead of invariant curves, some
invariant sets of complicated structure appear, that apparently
have non-zero measure. (The structure of one of these sets under
magnification is shown in {Fig. 4}; for similar structure for
discontinuous map see \cite{ashw} and references there). Numerics
show that, similarly to the smooth case, in this singular case
there are several regions  (see Figs. 5 -- 6): 1) the region where
there are no visible islands; in this region apparently almost all
trajectories are chaotic and correspond to escape to infinity; 2)
the region where there are islands, there are phase points
corresponding to quite fast escape to infinity and there are also
phase points outside the islands that do not escape for very long
time ($10^6$ iterations of map $\hat{\Pi}$ do not lead to escape);
about these latter phase points it is not clear if they are
captured forever or not; 3) the region were there are island and
all phase points outside the islands do not escape for very long
time; again about these latter phase points it is not clear if
they are captured forever or not.

One of islands inside the domain where lines $h=$ const are of the
type B is shown in  Fig. 7. For all initial points inside such an
island  the light shell on the period of its motion passes through
the heavy shell with non-zero relative velocity and the number of
passages of the shells through each other is the same. Therefore,
the Poincar\'e return map on such an island is an analytic map and
KAM-theory is applicable for description of dynamics inside this
island.

Numeric simulations give for the lower boundary of the domain
where there are no visible islands the scaling $h\sim
-\varepsilon^{0.38}$ (see Fig. 5).

\begin{figure*}
\includegraphics[width=.90\textwidth]{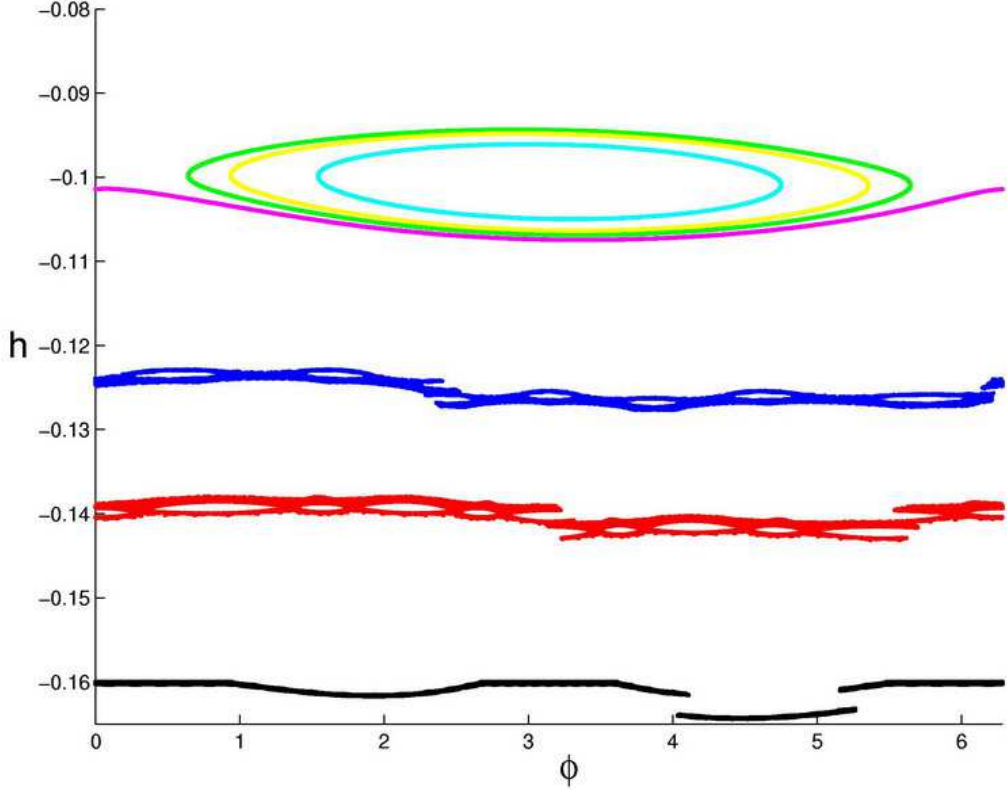}
\label{fig3} \caption{ Trajectories of seven initial points
 under the action of Poincar\'e return map in the domain
 where  lines $h=$const are of the type B and there are no
  visible escaping trajectories. $m_1/M=0.0003$, $J_1=1.4\sqrt{2}$,
  $J_2=1.2\sqrt{2}$, $H_1=-0.09$.}
\end{figure*}

\begin{figure*}
\includegraphics[width=1.0\textwidth,height=.50\textheight]{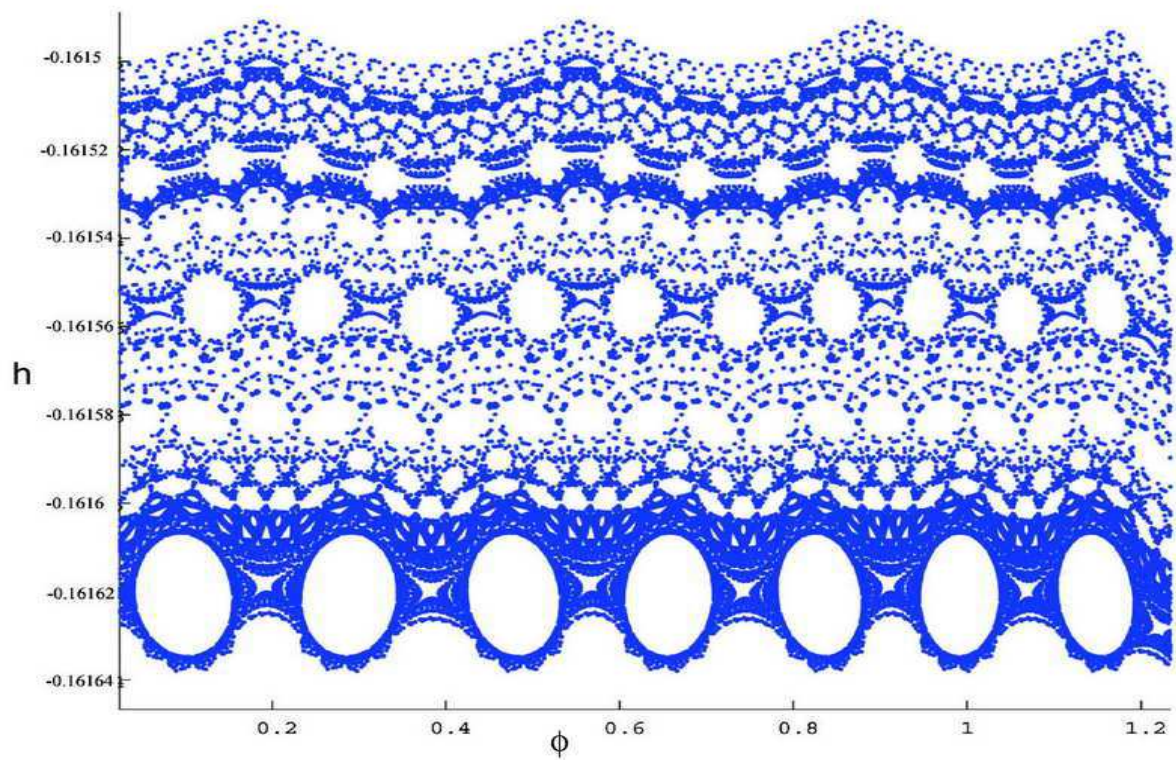}
\label{fig4} \caption{ The structure of invariant set from Fig. 3
under magnification. }
\end{figure*}

\begin{figure*}
\includegraphics[width=.90\textwidth]{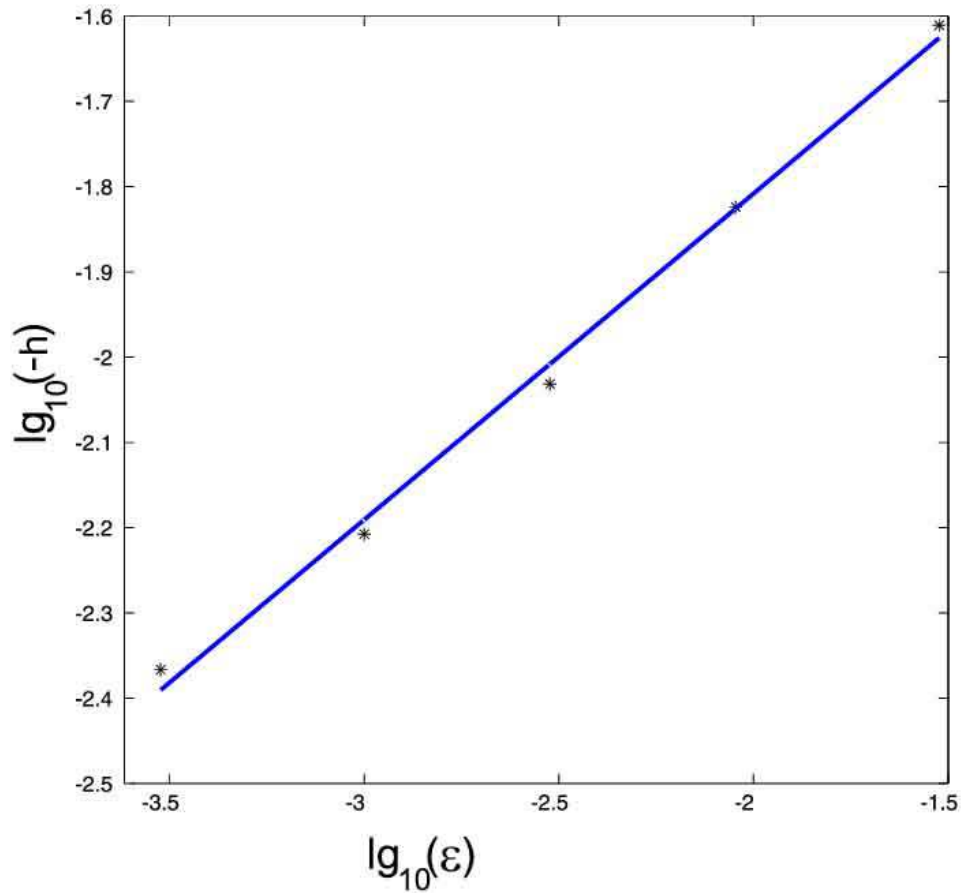}
\label{fig5} \caption{ Minimal values of $h$ such that there are
no visible islands at bigger values of $h$. The line $h=0$ is of
type B.}
\end{figure*}

\begin{figure*}
\includegraphics[width=.90\textwidth]{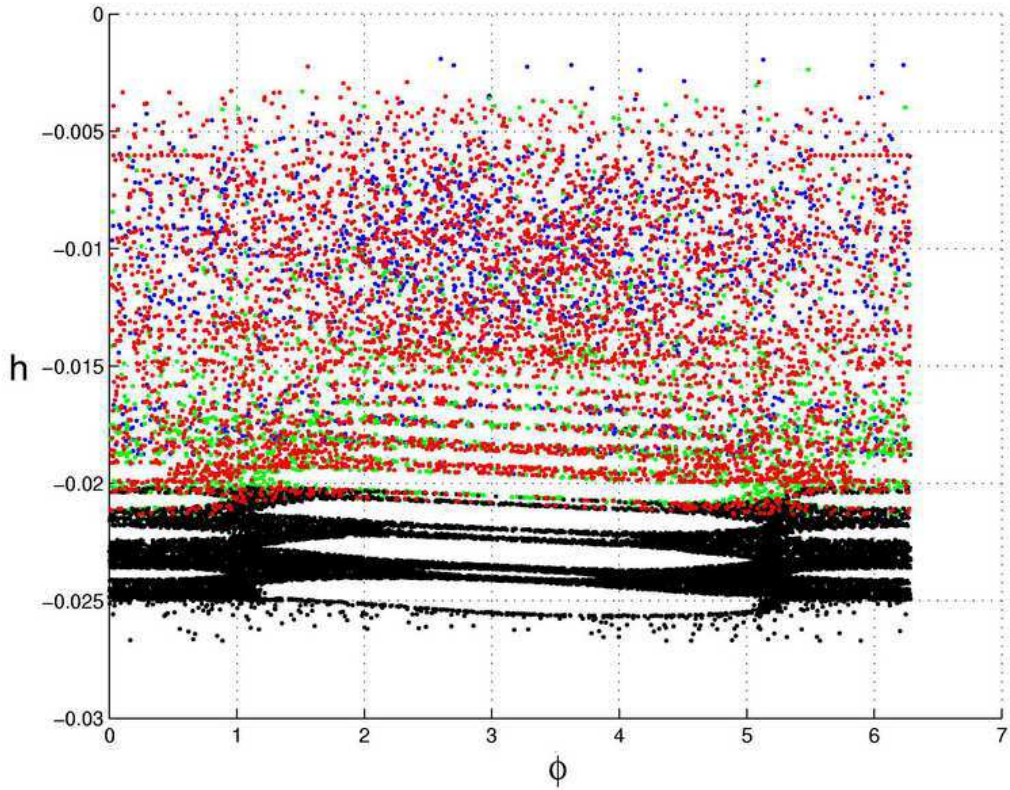}
\label{fig6}
 \caption{ Trajectories of four initial points
 under the action of Poincar\'e return map for the case
 when the line $h=0$ is of  type B.  $m_1/M=0.01$, $J_1=1.5\sqrt{2}$,
  $J_2=2.0\sqrt{2}$, $H_1=-0.09$.}
\end{figure*}

\begin{figure*}
\includegraphics[width=.90\textwidth]{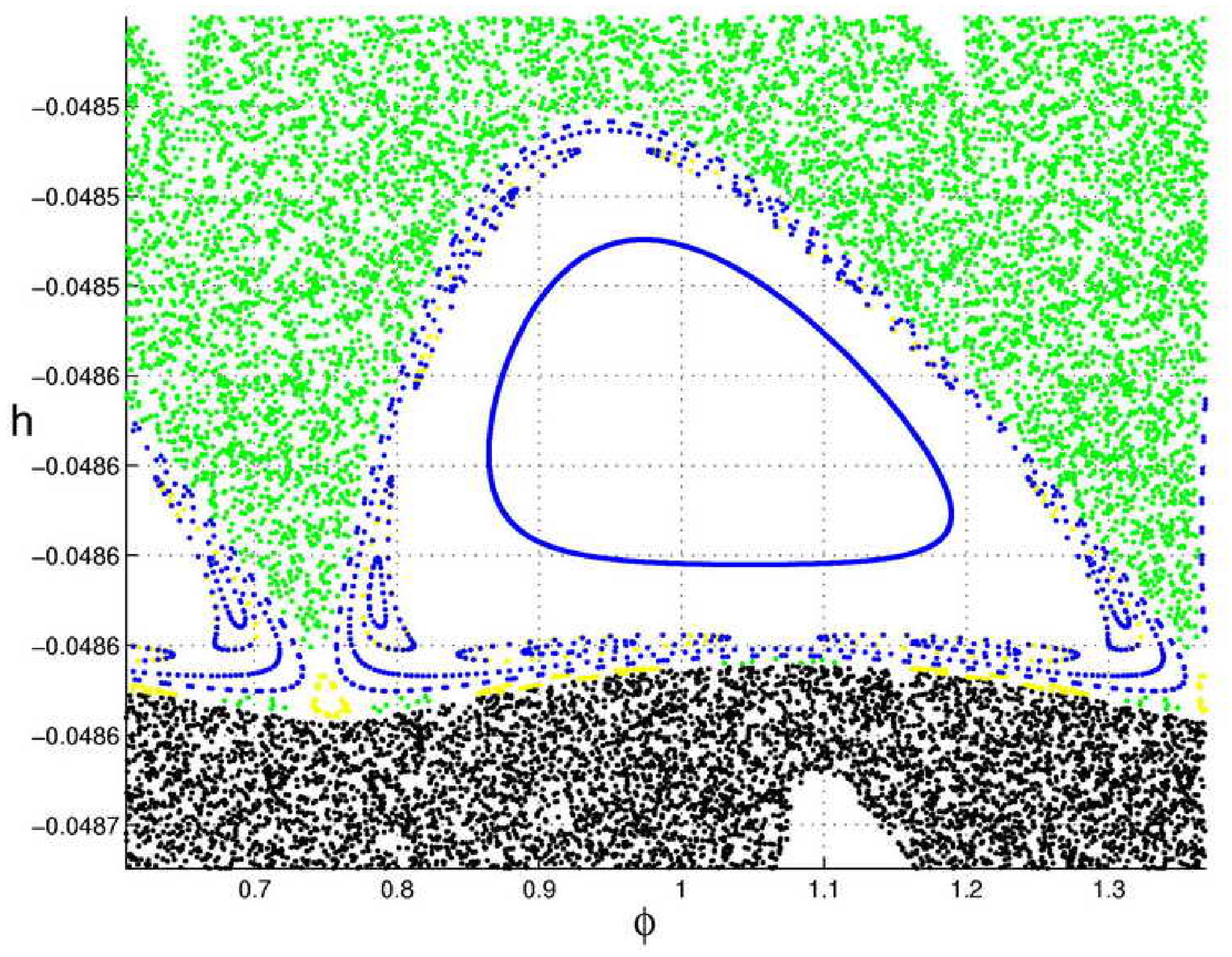}
\label{fig7}
 \caption{ Stability island inside the domain where
 lines $h=$ const are of type B.}
\end{figure*}

\section{ Dynamics of motion and Poincar\'e return map for
the non restricted problem.}

In the non restricted problem the motion of shells is described by
Hamiltonian $H=H_1+H_2$. In this problem $H$ is the total energy
of the system, and it is an integral of motion. Here
\begin{eqnarray}
 \label{ch7}
  H_1&=&\frac{p_1^2}{2m_1}+V_{1}, \nonumber \\
 V_1&= &
 \left\{
    \begin{array}{lll}
        V_{10},& \mbox{if} & r_1\leq r_2,\\
        V_{10}-{Gm_2m_1}/{r_1},&\mbox{if}& r_1 \ge r_2,
    \end{array}
 \right.
\end{eqnarray}
where
 $$
 V_{10}=-\frac{G(M+m_1/2)m_1}{r_1}+\frac{J_1^2m_1}{2r_1^2},
 $$
and
\begin{eqnarray}
 \label{ch8}
  H_2&=&\frac{p_2^2}{2m_2}+V_{2}, \nonumber \\
 V_2&= &
 \left\{
    \begin{array}{lll}
        V_{20},& \mbox{if} & r_2\leq r_1,\\
        V_{20}-{Gm_1m_2}/{r_2},&\mbox{if}& r_2 \ge r_1,
    \end{array}
 \right.
\end{eqnarray}
with
 $$
 V_{20}=-\frac{G(M+m_2/2)m_2}{r_2}+\frac{J_2^2m_2}{2r_2^2}.
 $$
Here $p_1$ is the momentum canonically conjugated to radius $r_1$,
and $p_2$ is the momentum canonically conjugated to radius $r_2$.
Therefore,  in the nonrestricted problem  motion of the shells is
described by a Hamiltonian system with two degrees of freedom.

\begin{figure*}
\includegraphics[width=.90\textwidth]{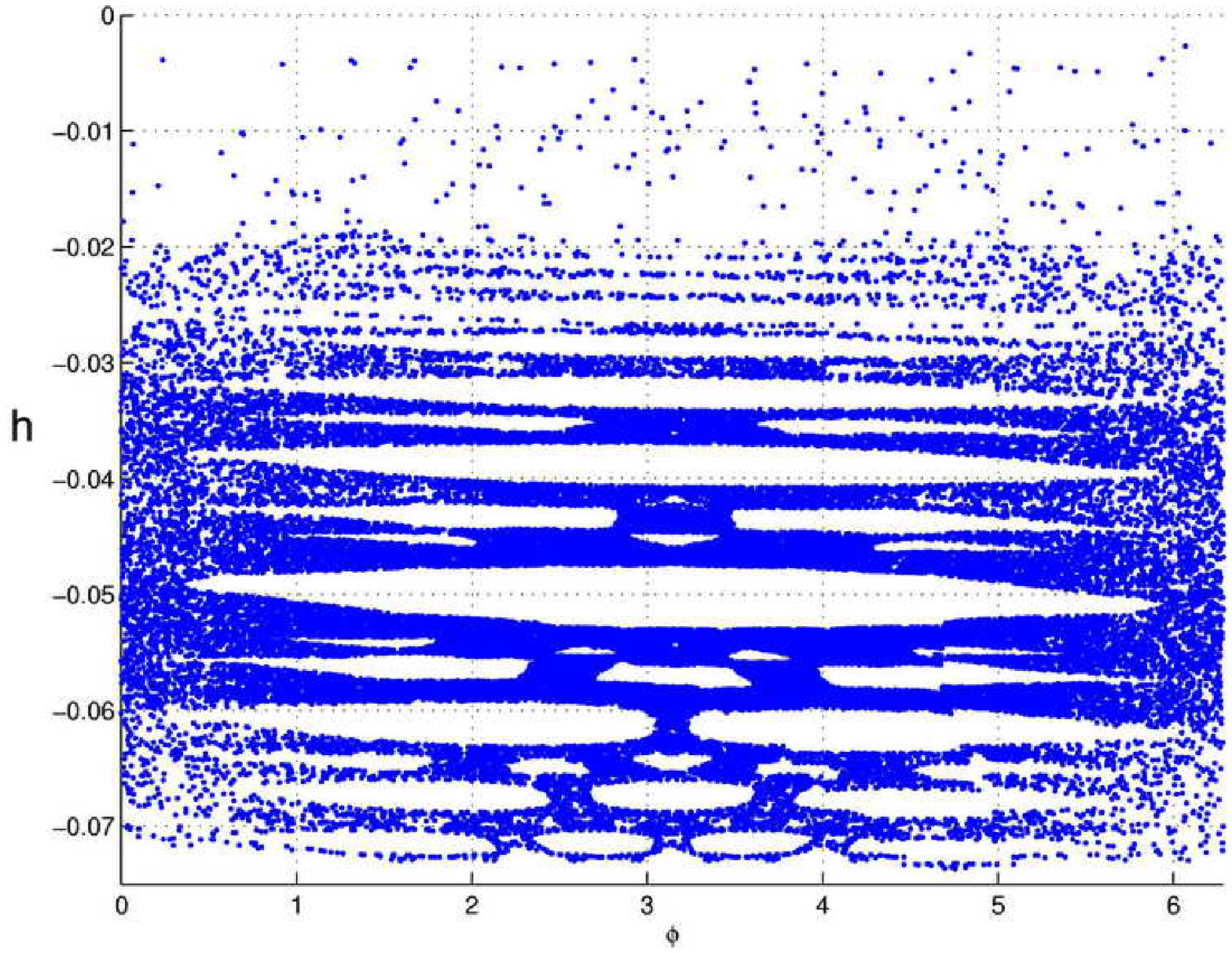}
 \caption{\label{fig8}  Trajectory of one initial point
 under the action of Poincar\'e return map for the nonrestricted
  problem. Parameters and initial conditions are: ${m_1=m_2=0.007}$,
 $J_1=1.5\sqrt{2},\; J_2=1.51\sqrt{2}$,  $H_1/m_1=-0.05$, $H_2/m_2=-0.05$,
 $r_1=5.1$,    $r_2=5.0$,
  the initial direction of the shells motion is inwards.}
\end{figure*}

We have constructed Poincar\'e return map for the non restricted
problem.  The total energy of system $H$ is fixed and the shells
with numbers "1"{} and "2"{} are now in equal rights. We have
constructed the Poincar\'e return map in the same way as in
section \ref{sec4}. At the time moments $t$, when the values of
$r_2$ have local minima on the trajectory, we  mark on the
cylinder $\Phi = \left\{ \left(h,\phi\right): -\infty <h<\infty,\;
\phi \mod 2\pi\right\}$ the values of the shell "2"{} energy
$H_2$, and the shell "1"{} mean anomaly $l_1 :
(H_2,l_1)=(h,\phi)$. We present one such trajectory for equal
masses $m_1=m_2=0.007$ in Fig \ref{fig8}. One see here chaotic
diffusion and a complicated structure of island.

\section{Conclusion.}

The shell approximation is often used for description of spherical
stellar clusters dynamics. We have considered dynamics of two
shells in the case when one can neglect the influence of  one
("light") shell onto the motion of another ("heavy") shell. It is
demonstrated that dynamics of  the light shell at small enough
absolute value of negative energy is chaotic and leads to escape
of the shell to infinity. There are two types of parameter values
for the system. For one type the Poincar\'e return map is analytic
in the domain of small negative energies of the light shell.
Therefore, KAM-theory is applicable for the description of the
dynamics provided that the mass of the heavy shell is much smaller
than the mass of the central body. In this case we have described
the structure of the phase space and have obtained the scaling
laws for the measure of the domain of chaotic motion and for the
minimal energy of the light shell sufficient for escape to the
infinity. For another  type of parameter values, the Poincar\'e
return map has singularities in the domain of small negative
energies of the light shell. In this case KAM-theory is not
applicable and it looks like currently there is no general theory
which would allow to deal with this situation. Numerical
calculation show coexistence of invariant sets of complicated
structure  and chaotic trajectories with final escape to infinity.

The work of G.S.B.-K. and M.V.B. was partly supported by RFBR
grant 99-02-18180, and INTAS grant 00-491. The work of A.I.N. was
partly supported by RFBR grant 03-01-00158, NSH-136.2003.1,
"Integration" B0053.

\end{document}